\documentclass[aps,prd,twocolumn,superscriptaddress]{revtex4}
\usepackage[hypertex]{hyperref}
\usepackage{graphicx} 
\usepackage{amsbsy}   
\usepackage{times}

\newcommand{\bvec}[1]{\boldsymbol{#1}}
\newcommand{\Expect}[1]{\left\langle{#1}\right\rangle}
\newcommand{\no}{\nonumber}

\usepackage{amsmath,amssymb,amscd,mathrsfs,fancybox,amsthm}
\usepackage{graphicx}
\usepackage{dcolumn}
\usepackage{bm}
\usepackage{bbold}

\begin{document}

\preprint{\normalsize UT-Komaba/10-3}

\title{Reflection Positivity of Free Overlap Fermions}

\author{Yoshio Kikukawa}
\affiliation{%
Institute of Physics, the University of Tokyo, Tokyo 153-8902, Japan
}%

\author{Kouta Usui}%
\affiliation{%
Department of Physics, the University of Tokyo, Tokyo 113-0033, Japan}
\affiliation{
Institute for the Physics and Mathematics of the Universe (IPMU),   the University of Tokyo, Chiba 277-8568, Japan
}%

\date{\today}

\begin{abstract}

It is shown that free lattice fermions defined by overlap Dirac operator  
fulfill the  Osterwalder-Schrader reflection positivity condition with respect to the link-reflection.  
The proof holds true in non-gauge models with interactions such as chiral Yukawa models. 

\end{abstract}

\pacs{11.15.Ha}
\maketitle

\noindent
\section{Introduction} 
%
To give the 
constructive definition of a relativistic quantum field theory through a lattice model, 
it is desirable for the lattice model to 
satisfy several fundamental requirements such as
locality, reflection positivity and hypercubic rotation and translation symmetry. 
%
Locality is believed to assure that
the continuum limit of the lattice model is universal
and the model belongs to the same universality class of the target continuum local field theory. 
Reflection positivity garantees that  the lattice model is a consistent quantum mechanical system satisfying unitarity.
Hypercubic rotation and translation symmetry is hopefully expected to result in the recovery of the Euclidean group symmetry in the continuum limit. 
These properties of the lattice model should be established rigorously, if possible,  before the applications to ``first-principle'' computation. 

In the lattice model, it is also desirable 
to keep the important symmetries of the target continuum field theory. 
Gauge symmetry can be preserved by introducing link variables. 
Chiral symmetry, despite of the infamous 
no-go theorem \cite{Karsten:1980wd,Nielsen:1980rz, Nielsen:1981xu}, can be also preserved by adopting lattice Dirac operators which satisfy the Ginsparg-Wilson (GW) relation \cite{Ginsparg:1981bj, Neuberger:1997fp, Neuberger:1998wv, Hasenfratz:1998ri, Hasenfratz:1998jp, Luscher:1998pqa}. 
For the overlap Dirac operator \cite{Neuberger:1997fp, Neuberger:1998wv}, 
a gauge-covariant solution to the GW relation which has been derived in the five-dimensional domain wall approach \cite{Kaplan:1992bt, Shamir:1993zy, Furman:1994ky},  a rigorous proof of locality has been given under a certain condition
for admissible gauge-link variables \cite{Hernandez:1998et}.  Even the applications to large scale numerical simulation of lattice QCD have been attempted, obtaining  clear numerical evidences for the spontaneous chiral symmetry breaking in QCD \cite{Fukaya:2007fb, Fukaya:2009fh, Hashimoto:2008fc}.  

However, for the GW fermions, reflection positivity 
is not fully understood yet \cite{Luscher:2000hn, Creutz:2004ir, Mandula:2009yd}. The situation should be compared to the case of Wilson fermions, for which  the rigorous proofs of reflection positivity have been given \cite{Osterwalder:1977pc, Luscher:1976ms, Menotti:1987cq}.
Then,  one might think that it is still somewhat premature to refer the above numerical applications as ``first-principle'' computations\footnote{Strictly speaking, locality is not also established for the gauge action adopted in these works and the phase structure of the gauge model is not fully understood.   See the arguments in \cite{Golterman:2003qe}. 
It would be desirable to clarify these points.}.

In this letter,  we will examine the reflection positivity of lattice fermions defined through 
overlap Dirac operator. 
It will be shown rigorously that free overlap Dirac fermions fulfill the reflection positivity with respect to the link-reflection.  
The proof will be extended to the cases of Majorana and Weyl fermions.
In ref \cite{Luscher:2000hn}, L\"{u}scher discussed the unitarity property of free overlap Dirac fermion
by investigating the positivity through the spectral representation of free propagator and concluded that 
free overlap Dirac fermion has a good unitarity property. 
Our direct proof of the reflection positivity given here is consistent with this observation.
Our proof 
will be also 
extended to the non-gauge models with interactions such as 
chiral Yukawa models.  
For gauge models, however, a proof of reflection positivity, if any, 
seems to be more involved and we will leave it for future study.

\vspace{.5em}
\noindent
\section{Reflection positivity} 
Reflection positivity  
is a sufficient condition for 
reconstructing a quantum theory in the canonical formalism, i.e. 
the Hilbert space of state vectors and 
the Hermitian Hamiltonian operator acting on the state vectors, 
from the lattice model defined in the Euclidean space\footnote{
Reflection positivity was first proposed in the continuum theory, 
in analyzing the relation between the
Minkowski and Euclidean versions of quantum field theory \cite{Osterwalder:1973dx}\cite{Osterwalder:1974tc}. 
}.
Let us formulate the reflection positivity condition for lattice Dirac fermions.
The cases of   Majorana and Weyl fermions will be discussed later. 

%
We assume a finite lattice
$\Lambda=[-L+1,L]^4\subset \mathbb{Z}^4$ in the lattice unit $a=1$, and 
impose \textit{anti}-periodic boundary condition in the time direction, 
and periodic boundary conditions in the space directions. 
The fermionic action is 
defined in the bilinear form
\begin{align}
\label{eq:Dirac-fermion-action}
A(\bar\psi,\psi)=\sum_{x\in\Lambda}\bar\psi(x)D_L\psi(x),
\end{align}
with  a lattice Dirac operator $D_L$\footnote{
The reader might prefer the sign convention where $A=-\bar\psi D_L \psi$ 
in stead of \eqref{eq:Dirac-fermion-action}.
These two sign convention in the fermionic action are connected to each other by the transformation
$\bar\psi'=i\bar\psi$ and $\psi'=i\psi$.
We have chosen this sign convention for the sake of the proof of the reflection positivity 
following \cite{Osterwalder:1977pc}. As we will see 
in section 3, 
this sign convention is suitable to prove the statement (iv) $\Delta A\in\mathcal{\bar P}$. 
}.
The kernel of the Dirac operator should be written as 
\begin{align}
D_L(x,y)=\sum_{n\in\mathbb{Z}^4}(-1)^{n_0}D(x+2nL,y),\quad x,y\in\Lambda,\label{D_L def}
\end{align}
where $D(x,y)$ is the kernel of the Dirac operator in the infinite lattice $\mathbb{Z}^4$. The quantum theory is then
completely characterized by the expectational functional defined by the fermionic path-integration:
\begin{align}
\label{eq:Expect-Dirac}
\Expect{F}:=\frac{1}{Z}\int{\cal D}[\psi]{\cal D}[ \bar \psi] \, {\rm e}^{A(\bar\psi,\psi)} F(\bar\psi,\psi), 
\end{align}
where the Grassmann integration for each field variable is specified as
$\int d \psi_\alpha(x) \psi_\alpha(x) =1$,  $\int d \bar\psi_\alpha(x) \bar\psi_\alpha(x) =1$, 
and the functional measure is defined by 
\begin{align} 
\label{eq:Dirac-fermion-measure}
{\cal D}[\psi]{\cal D}[ \bar \psi]  :=
\prod_{x \in \Lambda ; \alpha=1,2,3,4} \{ d \psi_\alpha(x)  d \bar \psi_\alpha(x)  \}. 
\end{align}

The reflection positivity condition --- a condition on this expectational functuonal --- is formulated as follows: let us define time reflection operator $\theta$
which acts on polynomials of the fermionic field variables by the relations
\begin{align}
\theta(\psi(x))&=\left( \bar\psi(\theta x)\gamma_0 \right)^T  \label{eq:t-reflection-psi}\\
\theta(\bar\psi(x))&=\left( \gamma_0\psi(\theta x) \right)^T \label{eq:t-reflection-bar-psi}\\
\theta(\alpha F +\beta G)&=\alpha^* \theta(F) +\beta^*\theta(G)\\
\theta(FG)&=\theta(G)\theta(F),
\end{align}
where we denote $\theta(t,\bvec x)=(-t+1,\bvec x)$ and $F,G$ are arbitrary polynomials of fermionic fields
and * means complex conjugation.
Let $\Lambda_\pm\subset\Lambda$ be the sets of sites with positive or non-positive time respectively.
Let $\mathcal{A_\pm}$ be the algebra of all the polynomials of the fields on $\Lambda_\pm$, and $\mathcal{A}$ 
on $\Lambda$. 
Then one says the theory is reflection positive if 
its expectation $\Expect\cdot:\mathcal{A}\to\mathbb{C}$ satisfies 
\begin{align}
\Expect{\theta(F_+)F_+}\ge0   \quad  \text{for} \, \,  {}^\forall F_+ \in \mathcal{A_+}  \label{RP}.
\end{align}

A popular choice of lattice Dirac operator is the Wilson Dirac operator, 
\begin{align}
D_{\rm w}
=
\sum_{\mu=0,1,2,3}\left\{ 
\frac{1}{2}\gamma_\mu(\partial_\mu-\partial_\mu^\dagger)+\frac{1}{2}\partial_\mu^\dagger\partial_\mu
\right\},
\end{align}
and in this case, the rigorous proofs of the reflection positivity have been given \cite{Osterwalder:1977pc, Luscher:1976ms, Menotti:1987cq}. The proofs cover the case with  gauge interaction. Therefore, 
the use of Wilson Dirac fermions in numerical applications has a completely sound basis\footnote{
One should note, however, that the reflection positivity is lost for the improved gauge and fermion actions. See \cite{Luscher:1984is} for the treatment for the case of the Symanzik improvement of Wilson gauge action.}.
Here we consider the overlap Dirac operator
\begin{align}
D&=\frac{1}{2}\Bigg(1+X\frac{1}{\sqrt{X^\dagger X}}\Bigg),\quad X=D_{\rm w}-m,
\end{align}
for $ 0 < m \le 1$.  This lattice Dirac operator  describes a single massless Dirac fermion and 
satisfies the GW relation, $\gamma_5 D + D \gamma_5 = 2 D \gamma_5 D$. Although 
the action is necessarily non ultra-local \cite{Horvath:1998cm}, 
the free overlap Dirac fermion indeed satisfies the reflection positivity condition, 
as will be shown below.\\


\noindent
\section{Proof of Reflection Positivity of overlap Dirac fermion}
%
%

To prove the reflection positivity, we need some additional 
definitions and notations.
First, let us denote 
\begin{align}
\Expect{F}_0:=  
 \int {\cal D}[\psi]{\cal D}[ \bar \psi] F(\bar\psi,\psi).
\end{align}
This $\Expect{\cdot}_0$ defines a linear function from ${\cal A}$ into $\mathbb{C}$.
Second, we decompose the lattice action $A$ into the following three parts :
\begin{align}
A=A_+ +A_- +\Delta A
\end{align}
where $A_+\in{\cal A_+}$, $A_-\in{\cal A_-}$, and $\Delta A$ is the part of the
action which contain both positive and negative time fields.
Thirdly, let us call $\mathcal{P}$ the set of all polynomials of the form 
$\sum_{j}\theta(F_{+ j})F_{+ j}$ in a finite summation, where $F_{+ j}\in\mathcal{A}_+$.

Although the above definition of ${\cal P}$ works well for the proof of the Wilson fermion, 
it is not enough for the proof of the overlap fermion. 
In our case of the overlap fermion, 
one needs to consider not only finite summations of the form $\sum_{j}\theta(F_{+ j})F_{+ j}$, 
but also infinite summations or integrations like 
\begin{align}\label{Riemanian sum}
\int ds\,\theta(F(s))F(s)=\lim_{N\to\infty}\sum_{k=1}^N \theta(F(s_k))F(s_k)\Delta s_k,
\end{align}
where the integration is defined as a limit of a finite Riemanian summation. 
 (see also eq. \eqref{Delta A positive} or \eqref{Delta A negative}).
%
To this end, we consider ${\cal \bar P}$, the closure of ${\cal P}$. The closure ${\cal \bar P}$
contains not only elements of the original ${\cal P}$, but also all the limit points of conversing sequences in ${\cal P}$.
That is, 
\begin{align}
F\in{\cal\bar P}  \quad \Leftrightarrow \quad  {}^\exists \{ F_n \}_{n=1}^\infty \in{\cal P} \, : \, \lim_{n\to\infty}F_n=F.
\end{align}
Here, the sequence $\{F_n\}_n \subset{\cal A}$ is defined to be convergent to some $F\in{\cal A}$, if 
any coefficient in $F_n$ converges to the corresponding coefficient in $F$ as a complex number
\footnote{
This definition of convergence in ${\cal A}$ is equivalent to the norm convergence of the Grassmann
algebra induced from the metric of the underlying vector space, where the fermionic fields
form an orthonomal basis.
}.
Note that with respect to this definition of convergence, 
the linear operation, the product operation
in $\mathcal{A}$, and the linear mappings $\Expect\cdot_0,\Expect\cdot:\mathcal{A}\to\mathbb{C}$ 
are all continuous functions, i.e.
if $F_n\to F$, $G_n\to G$, then 
\begin{align}
\alpha F_n +\beta G_n\to \alpha F +\beta G, \quad F_n G_n\to FG, \label{conti of prod and linear combi}\\
\Expect{F_n}\to\Expect{F},\quad \Expect{F_n}_0\to\Expect{F}\label{conti of expectations}.
\end{align}
%

Now, 
we note the fact  that the following four statements (i)-(iv) imply the reflection positivity: 

\vspace{.5em}
\noindent
(i) If $F,G$ belong to ${\cal\bar P}$ then $FG$ also belongs to ${\cal\bar P}$.  \\
(ii) For all $F\in{\cal\bar P}$, $\Expect{F}_0\ge0$.\\
(iii) $\theta(A_+)=A_-$.\\
(iv) $\Delta A\in{\cal \bar P}$.

\vspace{.5em}
\noindent
In fact, from these statements, it follows that 
\begin{align}\label{zero positivity}
\Expect {{\rm e}^A\,\theta(F_+)F_+ }_0 &=\Expect {{\rm e}^{A_+ +A_- +\Delta A}\,\theta(F_+)F_+ }_0 \no\\
&=\Expect {{\rm e}^{A_+ +\theta(A_+) +\Delta A}\,\theta(F_+)F_+ }_0 \no\\
&=\Expect {\underbrace{\theta({\rm e}^{A_+}){\rm e}^{A_+}\,{\rm e}^{\Delta A}}_{\in\mathcal{\bar P}\;\text{(by (i),(iv))}}
\underbrace{\theta(F_+)F_+}_{\in\mathcal{\bar P}} }_0\ge 0  
\end{align}
 for arbitrary $F_+\in\mathcal{A_+}$. Considering the special case where $F_+=1\in\mathcal{A}_+$, we have
$\Expect{{\rm e}^A}_0\ge 0$. Hence, we obtain
\begin{align}\label{zero to general}
\Expect {\theta(F_+)F_+}&= \frac{\Expect {{\rm e}^A\,\theta(F_+)F_+ }_0}{\Expect{{\rm e}^A}_0}\ge 0.
\end{align}
Therefore 
the proof is reduced to showing these four statements (i)-(iv).

Next, we will give the proofs of the statements (i)-(iv).
The statement (i) follows from the similar statement with ${\cal P}$, 
which has been proved for the Wilson case \cite{Osterwalder:1977pc} . In fact, 
let $F,G \in\mathcal{\bar P}$. Then, there exist sequences $\{F_n\}_{n}$ and $\{ G_n \}_n$ in ${\cal P}$ such that 
\begin{align}
F_n\to F,\quad G_n\to G.
\end{align}
From the continuity of the product operation in $\mathcal{A}$
(see \eqref{conti of prod and linear combi}), we get
\begin{align}
F_nG_n\to FG.
\end{align}
Since $F_n,G_n\in\mathcal{P}$, $F_nG_n\in\mathcal{P}$. Therefore $FG$ is the limit of the sequence $\{F_nG_n\}_n\subset \mathcal{P}$,
which means that $FG\in\mathcal{\bar P}$.

To show the statement (ii), one should refer to the definition of fermionic integration measure. 
With the definition (\ref{eq:Dirac-fermion-measure}),  it is sufficient to consider  
$F_+\in\mathcal{A}_+$ of the form
\begin{align}
F_+ = \prod_{x\in\Lambda_+ ; \alpha=1,2,3,4} \{ \bar \psi_\alpha(x)  \psi_\alpha(x)   \} \in \mathcal{P},  
\end{align}
for which one can see  
\begin{align}
\int {\cal D}[\psi]{\cal D}[ \bar \psi] \,  \theta(F_+) F_+ = \{ \det (\gamma_0^2) \}^{16L^4}= 1\ge 0.
\end{align}
Therefore, one concludes that for arbitrary $F\in\mathcal{P}$, $\Expect{F}_0\ge 0$.
Take arbitrary $F\in\mathcal{\bar P}$. Then there exists a converging sequence $\{F_n\}_n$ such that $F_n\to F$.
From the continuity of $\Expect{\cdot}_0$ (see \eqref{conti of expectations}), we obtain
\begin{align}
\Expect{F}_0=\Expect{\lim_{n\to\infty}F_n}_0=\lim_{n\to\infty}\Expect{F_n}_0\ge 0.
\end{align}

 The statement (iii) can be shown by using the property of 
the overlap Dirac kernel: $D_L^\dagger(x,y)=\gamma_0D_L(\theta x,\theta y)
\gamma_0$. In fact, one gets
\begin{align}
	\theta(A_+) &=\sum_{x\in\Lambda_+}\sum_{y\in\Lambda_+}\theta\Big(\bar\psi(x)D_L(x,y)\psi(y)\Big) \no\\
	&=\sum_{x\in\Lambda_+}\sum_{y\in\Lambda_+}\bar\psi(\theta y)\underbrace{\gamma_0D_L^\dagger (y,x)\gamma_0}_{
	=D(\theta y, \theta x)}\psi(\theta x) \no\\
	&=\sum_{x'\in\Lambda_-}\sum_{y'\in\Lambda_-}\bar\psi(x')D_L(x',y')\psi(y') \no\\
	&=A_-.
	\end{align}

\begin{figure}[b]
\begin{center}
\begin{picture}(250,120)
\put(0,60){\vector(1,0){100}}
\put(50,10){\vector(0,1){105}}
\thicklines
\put(20,60){\line(1,0){60}}
\thicklines
\put(20,60){\line(1,0){60}}
\thicklines
\put(80,60){\vector(0,1){25}}
\thicklines
\put(80,85){\line(0,1){25}}
\thicklines
\put(53,110){\line(1,0){27}}
\thicklines
\put(53,110){\line(0,-1){40}}
\thicklines
\put(53,70){\line(-1,0){6}}
\thicklines
\put(47,70){\line(0,1){40}}
\thicklines
\put(47,110){\line(-1,0){27}}
\thicklines
\put(20,110){\vector(0,-1){25}}
\thicklines
\put(20,85){\line(0,-1){25}}

\put(43,50){0}
\put(75,50){$\pi$}
\put(13,50){$-\pi$}
\put(56,70){$iE_1$}
\put(50,70){\circle*{3}}
\put(102,57){Re}
\put(46,118){Im}
\put(83,105){$\uparrow\infty$}

\thinlines
\put(120,60){\vector(1,0){100}}
\put(170,5){\vector(0,1){105}}
\thicklines
\put(140,60){\line(1,0){60}}
\thicklines
\put(140,60){\line(1,0){60}}
\thicklines
\put(200,60){\vector(0,-1){25}}
\thicklines
\put(200,35){\line(0,-1){25}}
\thicklines
\put(173,10){\line(1,0){27}}
\thicklines
\put(173,10){\line(0,1){40}}
\thicklines
\put(173,50){\line(-1,0){6}}
\thicklines
\put(167,50){\line(0,-1){40}}
\thicklines
\put(167,10){\line(-1,0){27}}
\thicklines
\put(140,10){\vector(0,1){25}}
\thicklines
\put(140,35){\line(0,1){25}}

\put(163,63){0}
\put(195,63){$\pi$}
\put(133,63){$-\pi$}
\put(176,45){$-iE_1$}
\put(170,50){\circle*{3}}
\put(222,57){Re}
\put(166,113){Im}
\put(203,10){$\downarrow-\infty$}
\end{picture}
\caption{Complex integration contours}
\label{fig:p0-integration-contour}
\end{center}
\end{figure}
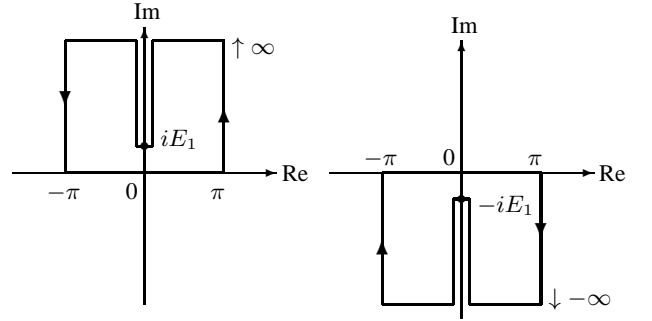

To show the statement (iv) $\Delta A\in\bar{\mathcal{P}}$, 
we use a spectral representation of $D_L(x,y)$.
To derive the spectral representation of $D_L$, we first Fourier transform 
the  overlap Dirac operator kernel $D(x,y)$ in the infinite volume:
\begin{align}
\label{eq-FT of D}
D(x,y)\Big|_{x_0\not=y_0}=\int\frac{d^4\bvec p}{(2\pi)^4}\,{\rm e}^{ip\cdot(x-y)}\frac{X(p_0,\bvec p)}{2\sqrt{X^\dagger X(p_0,\bvec p)}},
\end{align}
where $X(p_0,\bvec p)=\sum_{\mu}i\gamma_\mu \sin p_\mu +\sum_\mu(1-\cos p_\mu)-m$.
Then, we change the $p_0$ integration region, $[-\pi,\pi]$, to the contours along the imaginary axis
in the complex $p_0$ plane by Cauchy's integration theorem, as shown in 
FIG.~\ref{fig:p0-integration-contour}. 
Depending whether $x_0 - y_0 > 0$ or  $x_0 - y_0 < 0$, 
we choose the contours 
$[i E_1, i \infty]$ or  $[-i E_1,- i \infty]$, respectively, to obtain
\begin{align}
&D(x,y)\Big|_{x_0-y_0>0}\no\\
&=\int\frac{d^3\bvec p}{(2\pi)^3}\int_{E_1}^\infty\frac{dE}{2\pi}\,{\rm e}^{-E(x_0-y_0)} {\rm e}^{i\bvec p\cdot (\bvec x-\bvec y)}
\frac{X(iE,\bvec p)}{\sqrt{-X^\dagger X(iE,\bvec p)}}\label{D positive}\\
&D(x,y)\Big|_{x_0-y_0<0}\no\\
&\quad=\int\frac{d^3\bvec p}{(2\pi)^3}\int_{E_1}^\infty\frac{dE}{2\pi}\,{\rm e}^{E(x_0-y_0)} {\rm e}^{i\bvec p\cdot (\bvec x-\bvec y)}
\frac{X(-iE,\bvec p)}{\sqrt{-X^\dagger X(iE,\bvec p)}}.\label{D negative}
\end{align}
where $E_1$ is the edge of the cut coming from the square root, and is determined by the relations
\begin{align}
X^\dagger X(iE_1,\bvec p)=0, \quad E_1>0.
\end{align}

In this spectrum representaion of $D$, 
it is very crucial to notice the fact that $\mp \gamma_0 X(\pm iE,\bvec p)$ ($E\ge E_1$) are positive definite matrices  
and there exist matrices $Y_{\pm}(E,\bvec p)$ such that
\begin{align}
\label{gamma0-X-positive}
\mp \gamma_0 X(\pm iE,\bvec p)=Y_{\pm}^\dagger Y_{\pm}(E,\bvec p) \quad (E\ge E_1). 
\end{align}
In fact, it is not difficult to check that $Y_{\pm}(E,\bvec p)$ are given by
\begin{align}
Y_{\pm}(E,\bvec p)=-\sum_{k=1}^3\frac{l(E,\bvec p)\sin p_k}{W(E,\bvec p)}\gamma_k \mp i\frac{W(E,\bvec p)}{2l(E,\bvec p)}\gamma_0+il(E,\bvec p), \no
\end{align}
where $W(E,\bvec p)=\sum_{k=1}^3(1-\cos p_k) +1 -\cosh E -m$ and 
\begin{align}
l(E,\bvec p)&= \left[ \frac{1}{2}\frac{\sinh E}{\sum_{k=1}^3\sin^2 p_k/W(E,\bvec p)^2+1}\times \right. \no\\
&\left. \qquad
\Bigg(1+ \sqrt{1-\frac{\sum_{k=1}^3\sin^2 p_k+W(E,\bvec p)^2}{\sinh^2 E}}\Bigg)\right]^{\frac{1}{2}}. 
\no
\end{align} 

From the equations \eqref{D_L def}, \eqref{D positive} and \eqref{D negative}, 
we find the spectrum representation of $D_L(x,y)$ as follows: putting $V=1/(2L)^3$, 
\begin{align}
D_L(x,y)\Big|_{x_0\not=y_0}&=\sum_{\bvec p}\int_{E_1}^\infty\frac{dE}{2\pi}\frac{1}{1+{\rm e}^{-2EL}}\frac{1}{V}\no\\
&\quad\times {\rm e}^{-E|x_0-y_0|}{\rm e}^{i\bvec p\cdot (\bvec x-\bvec y)}
\frac{X(\epsilon iE,\bvec p)}{\sqrt{-X^\dagger X(iE,\bvec p)}}\no\\
&+\sum_{\bvec p}\int_{E_1}^\infty\frac{dE}{2\pi}\frac{{\rm e}^{-2EL}}{1+{\rm e}^{-2EL}}\frac{1}{V}\no\\
&\quad\times {\rm e}^{E|x_0-y_0|}{\rm e}^{i\bvec p\cdot (\bvec x-\bvec y)}
\frac{-X(-\epsilon iE,\bvec p)}{\sqrt{-X^\dagger X(iE,\bvec p)}},\label{D_L}
\end{align}
where $\epsilon$ is defined as the sign of $x_0-y_0$, and the spacial momentum $p_k$ runs over $p_k=n_k\pi/L,\,(-L\le n\le L)$
in the above summation. 
In \eqref{D_L}, 
the first term  becomes $D(x,y)$ in the limit $L\to\infty$, and
the second term represents a `finite lattice effect' which vanishes in the limit $L\to\infty$. 
The latter is the contribution of the  wrong-sign-energy modes and 
the minus sign appearing in front of $X(-\epsilon iE,\bvec p)$
comes from the \textit{anti}-periodicity in the time direction, which  
is required 
for the positivity,  as will be seen. 



From these observations, now we can show that $\Delta A\in\bar{\mathcal{P}}$: 
for the term with $x_0>0$, $y_0\le 0$ (in this case $\epsilon=1$), we obtain
\begin{align}
&\sum_{x\in\Lambda_+}\sum_{y\in\Lambda_-}\bar\psi(x)D_L(x,y)\psi(y)\no\\
&=-\sum_{\bvec p}\int_{E_1}^\infty\frac{dE}{2\pi}\frac{1}{V}\Bigg[C_{E,\bvec p}\theta(C_{E,\bvec p})+D_{E,\bvec p}\theta(D_{E,\bvec p})\Bigg],\label{Delta A positive}
\end{align}
where $C_{E,\bvec p}$ and $D_{E,\bvec p}$ are defined by
\begin{align}
&C_{E,\bvec p}\no\\
&=\sqrt{\frac{1}{1+{\rm e}^{-2EL}}}\sum_{x\in\Lambda_+}\bar\psi(x)\gamma_0\tilde{Y_+}(E,\bvec p)^\dagger 
{\rm e}^{-Ex_0}{\rm e}^{i\bvec p\cdot\bvec x},\\
&D_{E,\bvec p}\no\\
&=\sqrt{\frac{{\rm e}^{-2EL}}{1+{\rm e}^{-2EL}}}\sum_{x\in\Lambda_+}\bar\psi(x)\gamma_0\tilde{Y_-}(E,\bvec p)^\dagger 
{\rm e}^{Ex_0}{\rm e}^{i\bvec p\cdot\bvec x},
\end{align}
with
$\tilde{Y_+}(E,\bvec p)=Y_+(E,\bvec p)/(-X^\dagger X(iE,\bvec p))^{\frac{1}{4}}$.
The overall minus sign in the r.h.s. of \eqref{Delta A positive} results 
from \eqref{D_L} by using \eqref{gamma0-X-positive}. 
This minus sign is canceled after exchanging
the order of the Grassmann products 
in \eqref{Delta A positive}, 
and we see that this term belongs to $\bar{\mathcal{P}}$.
%
Similarly, for the term with $x_0\le0$, $y_0>0$ (in this case $\epsilon=-1$), we obtain
\begin{align}
&\sum_{x\in\Lambda_-}\sum_{y\in\Lambda_+}\bar\psi(x)D_L(x,y)\psi(y)\no\\
&=\sum_{\bvec p}\int_{E_1}^\infty\frac{dE}{2\pi}\frac{1}{V}
\Bigg[\theta(C'_{E,\bvec p})C'_{E,\bvec p}+\theta(D'_{E,\bvec p})D'_{E,\bvec p}\Bigg],\label{Delta A negative}
\end{align}
where $C'_{E,\bvec p}$ and $D'_{E,\bvec p}$ are defined by
\begin{align}
&C'_{E,\bvec p}\no\\
&=\sqrt{\frac{1}{1+{\rm e}^{-2EL}}}\sum_{y\in\Lambda_+} \tilde{Y_-}(E,\bvec p) \psi(y) {\rm e}^{-E y_0} {\rm e}^{-i\bvec p\cdot\bvec y},\\
&D'_{E,\bvec p}\no\\
&=\sqrt{\frac{{\rm e}^{-2EL}}{1+{\rm e}^{-2EL}}}\sum_{y\in\Lambda_+}\tilde{Y_+}(E,\bvec p) \psi(y) {\rm e}^{Ey_0}{\rm e}^{-i\bvec p\cdot\bvec y}. 
\end{align}
In this case we do not need to exchange the order of the product,
and we immediately see that this term belongs to $\bar{\mathcal{P}}$.
Thus we obtain $\Delta A\in\bar{\mathcal{P}}$ and complete the proof of the reflection positivity in the Dirac case. 

\vspace{.5em}
\noindent
\section{Majorana and Weyl fermions}
The above  proof for Dirac fermion can be extended for 
Majorana fermions\cite{Neuberger:1997bg, Kaplan:1999jn}
and 
Weyl fermions\cite{Narayanan:1993sk, Narayanan:1993ss, Narayanan:1994gw, Luscher:1999mt}. 
In the case of Majorana fermion, 
the anti-field $\bar \psi$ is the charge conjugation of the field $\psi$ : 
$\bar \psi \equiv \psi^T C$, where $C$ is  the charge conjugation matrix satisfying
$C\gamma_\mu C^{-1}=-\gamma_\mu^T$, $C\gamma_5 C^{-1}=\gamma_5^T$, 
$C^\dagger C=1,C^T=-C$. 
Accordingly, the path-integral measure reduces to 
${\cal D}[\psi] := s 
\prod_{x \in \Lambda; \alpha} \{ d \psi_\alpha(x)  \}$, where
a sign factor $s (=\pm 1)$ is introduced for later convenience. 
This Majorana-reduction does not contradict 
with the  definition of time reflection $\theta$, because both 
(\ref{eq:t-reflection-psi}) and (\ref{eq:t-reflection-bar-psi}) imply 
$\theta(\psi(x)) = C \gamma_0 \psi(\theta x)$. 
Then, 
the conditions (iii) $A_-=\theta(A_+)$ and (iv) $\Delta A\in\bar{\mathcal{P}}$ follow immediately. 
The propery (ii) of $\bar{\mathcal{P}}$ also holds ture 
by  the fact
that one can always choose the sign factor $s$ so that 
$\int {\cal D}[\psi] \theta(F_+) F_+ = \{ \det (C \gamma_0) \}^{8L^4} >0$
for $F_+ =\prod_{x\in\Lambda_+ ; \alpha=1,2,3,4}  \{ \psi_\alpha(x)  \}$. Thus the reflection positivity (\ref{RP}) follows from the conditions (i), (ii), (iii) and (iv) also for the Majorana case. 

For Weyl fermion, we define the chiral components by 
\begin{equation}
\psi_\pm(x) = \left(\frac{1\pm \hat\gamma_5}{2}\right) \psi(x), \quad 
\bar \psi_\pm(x) = \bar \psi(x)  \left(\frac{1\mp \gamma_5}{2}\right), 
\end{equation}
where $\hat \gamma_5 = \gamma_5(1-2D)$. 
We adopt, for simplicity,  the chiral basis for gamma matrices
in which $\gamma_5 = \sigma_3 \otimes \mathbb{1} $, 
$\gamma_0=\sigma_1 \otimes \mathbb{1}$, and denote the spinor indices as
$\alpha_+=\{1,2\}, \alpha_-=\{3,4\}$. Then $\bar\psi_{\pm \alpha_\mp}(x) = \bar \psi_{\alpha_\mp}(x)$.
The action of the Weyl fermion is given by $A^{(\pm)}= \sum_x \bar\psi_\pm (x) D_L \psi_\pm(x)$,
and the Dirac fermion action  (\ref{eq:Dirac-fermion-action}) decomposes as 
$A=A^{(+)}+A^{(-)}$.
To define  the path-integral  for the Weyl fermion, 
we introduce the chiral bases
\begin{equation}
\{ v^i_{\pm}(x) \, \vert \, \hat \gamma_5 v^i_\pm (x) = \pm v^i_\pm(x); \,  i=1, \cdots, n_\pm\}, 
\end{equation}
where $n_\pm= 2(2L)^4$, and expand the fields as $\psi_\pm(x) = \sum_i v^i_\pm (x) c^i_\pm$.  
The path-integral measure is then defined by  
\begin{equation}
{\cal D} [\psi_\pm ] {\cal D} [\bar \psi_\pm ] =
\prod_i  d c^i_\pm \prod_{x \in \Lambda; \alpha_\mp} d \bar \psi_{\alpha_\mp}(x) , 
\end{equation}
and the path-integral measure of Dirac fermion (\ref{eq:Dirac-fermion-measure}) is factorized as 
$ {\cal D}[\psi]{\cal D}[ \bar \psi] 
= J \, {\cal D} [\psi_+]{\cal D}[\bar \psi_+] {\cal D}[\psi_-]{\cal D}[\bar \psi_-]$ where 
the Jacobin 
\begin{eqnarray}
&& J[ \psi_\alpha(x); c_+^i, c_-^j] \nonumber\\
&&= \left\vert 
{\small
\begin{array}{cccccc}
v_{+\alpha}(x)^1 & \cdots & v_{+\alpha}(x)^{n_+} & v_{-\alpha}(x)^1 & \cdots & v_{-\alpha}(x)^{n_-}
\end{array}
}
\right\vert 
 \nonumber\\
\end{eqnarray}
can be set to unity by 
choosing the chiral basis vectors appropriately.
The expectational functional for the left-handed Weyl fermion  is defined by 
\begin{align}
\label{eq:Expect-Weyl}
\Expect{F}^{(-)}:=
\frac{1}{Z^{(-)}} \int{\cal D}[\psi_-]{\cal D}[ \bar \psi_-] \,
 {\rm e}^{A^{(-)}(\bar\psi_-,\psi_-)} F(\bar\psi_-,\psi_-). 
\end{align}
Because of the factorization properties of the action and the path-integral measure,  we note  the identity
\begin{eqnarray}
\Expect{F(\bar\psi_-,\psi_-)}^{(-)} &=& \Expect{F(\bar\psi_-,\psi_-)}^{(-)} \Expect{1}^{(+)}  \nonumber \\
&=&  \Expect{F(\bar\psi_-,\psi_-)} . 
\end{eqnarray}

In this setup, we can formulate the reflection positivity condition for the Weyl fermion as follows. 
We define the time reflection operator $\theta$  for the left-handed fields as 
\begin{align}
\theta(\psi_{- {\alpha_-} } (x))
&=\{ \bar\psi_-(\theta x)\gamma_0 \}_{\alpha_-} =  \bar\psi_{- \alpha_+}(\theta x), \\
\theta(\bar\psi_{- {\alpha_+}}(x))&=\{ \gamma_0\psi_-(\theta x) \}_{\alpha_+} = \psi_{- \alpha_-}(\theta x) , 
\end{align}
where $\alpha_+ = \alpha_- - 2$. 
And let $\mathcal{A_\pm}^{(-)}$ be the algebra of all the polynomials of the left-handed field components 
 $\psi_{- \alpha_-}(x)$ and $\bar\psi_{- \alpha_+}(x)$ on $\Lambda_\pm$. Then one can show
 \begin{align}
\Expect{\theta(F_+)F_+}^{(-)} \ge0   \quad  \text{for} \, \,  {}^\forall F_+ \in \mathcal{A_+^{(-)}}  \label{RP-Weyl}.
\end{align}
Note that, in this formulation of the reflection positivity,  the field components $\psi_{- \alpha_+}(x)$ are completely excluded from observables.
 
To prove (\ref{RP-Weyl}), we note the fact that 
the expectational functional for the left-handed Weyl fermion (\ref{eq:Expect-Weyl})
is simply related to the expectational functional for the Dirac fermion (\ref{eq:Expect-Dirac}) by
$\Expect{F} ^{(-)}=\Expect{F} ^{(-)} \Expect{1} ^{(+)} = \Expect{F} $ for $F(\bar\psi_-,\psi_-)$. Moreover, 
since
\begin{equation}
\left\{  \left(\frac{1-  \hat\gamma_5}{2}\right) D^{-1}\ \right\}_{\alpha_-, \alpha_+} = 
\left\{  \left(\frac{1- \gamma_5}{2}\right) D^{-1} \right\}_{\alpha_-, \alpha_+} , 
 \end{equation}
one can show 
\begin{align}
\Expect{   F(\psi_{- \alpha_-}, \bar\psi_{- \alpha_+})} ^{(-)} = 
\Expect{   F(\psi_{\alpha_-}, \bar\psi_{\alpha_+})       } , 
\end{align}
by performing the  Wick contructions explicitly. 
Then, (\ref{RP-Weyl}) follows immediately from the reflection positivity condition (\ref{RP}) 
for the overlap Dirac fermion. 
\section{Reflection positivity of chiral Yukawa theory}
In this section, we consider the case with an interaction --- chiral Yukawa model.
Chiral Yukawa model is defineded by the action\cite{Luscher:1998pqa}
\begin{align}\label{Yukawa action}
A&=\sum_{x\in\Lambda}\Big\{\bar\psi D\psi -\phi^*\partial_\mu^\dagger\partial_\mu \phi\no
- m_0^2 \phi^* \phi - \frac{\lambda_0}{2}  ( \phi^* \phi )^2 -2\bar\chi \chi 
\\
&\qquad+g_0(\bar\psi+\bar\chi)
\big\{\frac{1}{2}(1-\gamma_5)\phi+\frac{1}{2}(1+\gamma_5)\phi^*\big\}
(\psi+\chi)\Big\}, 
\end{align}
where $\psi$ is a Dirac field, $\chi$ is an auxiliary Dirac field, and $\phi$ is a complex scalar field.
In this case, we define the field algebra $\mathcal{A}^{\rm(Y)}$ of the chiral Yukawa theory
 as the set of all the
polynomials $F(\psi,\chi,\phi)$ of fermionic fields $\psi$ and $\chi$
whose coefficients are complex valued continuous (not necessarily holomorphic) 
functions of bosonic field configuration $\phi$,
with converging expectation value $\Expect{F}^{\rm(Y)}$ defined through the path integration as usual:
\begin{align}\label{expect for yukawa}
\Expect{F}^{\rm(Y)}:=\frac{1}{Z^{\rm(Y)}}
\int\mathcal{D}[\text{path}]\,{\rm e}^{A(\psi,\chi,\phi)}F(\psi,\chi,\phi)<\infty.
\end{align}
Here, $\mathcal{D}[\text{path}]$ stands for  the path integration measure, 
\begin{align}
\mathcal{D}[\text{path}]= \mathcal{D}[\psi]\mathcal{D}[\bar\psi]\mathcal{D}[\chi]\mathcal{D}[\bar\chi]\mathcal{D}[\phi]\mathcal{D}[\phi^*].
\end{align} 
Note that all the polynomials of bosonic field configuration belong to $\mathcal{A}^{\rm (Y)}$.

The $\theta$ operation for the fermionic fields $\psi, \chi$ is the same as in the free case 
\eqref{eq:t-reflection-psi}\eqref{eq:t-reflection-bar-psi}. 
For the bosonic field $\phi$, the $\theta$ reflection is defined as
\begin{align}
\theta \phi(x):=\phi(\theta x).
\end{align}
For $F\in\mathcal{A}^{\rm(Y)}$ of the form $F(\psi, \chi, \phi)=f(\phi)M(\psi,\chi)$ with $f$ being a continuous 
function
 of $\{\phi(x)\}_{x\in\Lambda}$ and $M(\psi,\chi)$ some monomial of $\{\psi_\alpha(x),\bar\psi_\alpha(x),
\chi_\alpha(x),\bar \chi_\alpha(x)\}_{x\in\Lambda}$, we define 
\begin{align}
\theta(F)(\psi, \chi, \phi)=f^*(\theta\phi)M^\dagger(\theta\psi,\theta \chi),
\end{align}
where $M^\dagger$ means the monomial whose order of the Grassmann product
is reversed in the original $M$. We extend the $\theta$ operation for arbitrary $F \in \mathcal{A}^{\rm(Y)}$
 by anti-linearity. 
 Then, 
 the reflection positivity of this chiral Yukawa model is defined in the same way as in the free overlap fermion
 case :
 \begin{align}
 \Expect{\theta(F_+)F_+}^{\rm(Y)}\ge 0 ,\quad  \text{for }  \, {}^\forall F_+\in\mathcal{A}_+^{\rm(Y)}.\label{RP of Yukawa}
 \end{align}
 
%
We will prove 
the reflection positivity of the chiral Yukawa model in the same manner as in 
the free fermion case, based on the statements (i)-(iv).
%
 The statement (i) clearly holds true. 
 In the statement (ii), 
 we define
 the expectation $\Expect{\cdot}_{0}^{\rm(Y)}$ for the Yukawa model as
 \begin{align}\label{zero expect for yukawa}
\Expect{F}_{0}^{\rm(Y)}:= \int\mathcal{D}[\text{path}]\,
 F(\psi,\chi,\phi).
 \end{align}
For this definition to make sense, $F$ should be 
a special element  in $\mathcal{A}^{\rm(Y)}$
so that  the right hand side of (\ref{zero expect for yukawa}) is convergent. 
Let $\mathcal{B}$ be the subset of  $\mathcal{A}^{\rm(Y)}$ whose elements 
are integrable with respect to the above $\Expect\cdot_0$-measure.  
Note that 
if $F$ belongs to $\mathcal{A}^{\rm(Y)}$, ${\rm e}^{A}F$ belongs to $\mathcal{B}$
because of the rapidly decreasing property of the bosonic weight. 
Then, the statement (ii) should be rephrased as :
(ii) For all $F\in{\cal \bar P}\cap\mathcal{B}$, $\Expect{F}_0\ge0$.
To show this,
it is sufficient to consider $F_+\in\mathcal{A}_+^{\rm(Y)}$ of the form
\begin{align}
F_+=f(\phi_+)\prod_{x\in\Lambda_+;\alpha}\bar\chi_\alpha(x)\chi_\alpha(x)\bar\psi_\alpha(x)\psi_\alpha(x),
\end{align}
where $\phi_+=\{\phi(x)\}_{x\in\Lambda_+}$. For such an $F_+$, we obtain
\begin{align}\label{Yukawa zero positivity}
\Expect{\theta(F_+)F_+}_{0}^{\rm(Y)}=\left|\int\mathcal{D}[\phi_+]\mathcal{D}[\phi^*_+]\,f(\phi_+) \right|^2 \ge 0.
\end{align}
Therefore, for arbitrary $F\in\mathcal{P}\cap\mathcal{B}$ the statement (ii) holds. This result can be extended for 
$F\in\mathcal{\bar P}\cap\mathcal{B}$
 by a similar argument in the case of free fermion given above
 \footnote{
 To give a precise proof, we need to clarify what we mean by saying that a sequence of bosonic
 functions $\{f_n\}_{n=1}^\infty$ converges to some $f$. Here, we use uniform convergence, i.e.
 a sequence $\{f_n\}_n$ is defined to converge to some $f$ iff
\[ \lim_{n\to\infty}\sup_{\phi}|f_n(\phi)-f(\phi)|=0. \]
As is well known in the calculus, with this definition of convergence, 
the limit function $f$ is guaranteed to be continuous, and
the order of the integration and the limit operation can be exchanged, 
implying the continuity of $\Expect{\cdot}_0$ and $\Expect{\cdot}$. In particular,
if $\{f_n\}_n\subset\mathcal{B}$ converges to some $f$, $\Expect{f_n}_0$ also
converges to $\Expect{f}_0$, meaning that $f\in\mathcal{B}$. 
 }.
 The statement (iii) can be checked by noting that the interaction terms are strictly local 
 and belongs to either $A_+$ or $A_-$,  depending on the time coordinate, and they are 
 mapped to each other by  the $\theta$ transformation.
 To show the statement (iv), one should note that only the first two terms in the action \eqref{Yukawa action} 
 contribute to $\Delta A$. 
 The first fermionic part belongs to $\mathcal{\bar P}$ as shown in the above proof of the free Dirac fermion. 
 As to the second bosonic part, it is a well-known fact. 

From these statements (i)-(iv), the reflection positivity of the
chiral Yukawa model follows immediately. 
By \eqref{zero positivity} and the fact that ${\rm e}^A\theta(F_+)F_+\in\mathcal{B}$
for arbitrary $F_+\in\mathcal{A}^{\rm (Y)}_+$,
we obtain 
\begin{align}
\Expect{{\rm e}^A\theta(F_+)F_+}^{\rm (Y)}_0\ge 0,\quad \Expect{{\rm e}^A}^{\rm (Y)}_0=\Expect{{\rm e}^A\theta(1)1}^{\rm (Y)}_0 \ge 0.
\end{align}
This implies
\begin{align}\label{zero to general}
\Expect {\theta(F_+)F_+}^{\rm (Y)}&= \frac{\Expect {{\rm e}^A\,\theta(F_+)F_+ }^{\rm (Y)}_0}{\Expect{{\rm e}^A}^{\rm (Y)}_0}\ge 0,
\quad{}^\forall F \in\mathcal{A}_+^{\rm(Y)},
\end{align}
completing the proof of \eqref {RP of Yukawa}.

\vspace{.5em}
\noindent
\section{Discussions}  There is another route to the proof: it is through the connection 
to domain wall fermion\cite{Neuberger:1997bg, Kikukawa:1999sy}. 
Since domain wall fermion is 
defined by the five-dimensional Wilson fermion,  it fulfills the reflection positivity by itself.  
In the free case, fortunately, the positivity condition is also satisfied for 
the Pauli-Villars fields, 
%
%
where the Pauli-Villars fields are defined by a five-dimensional  Wilson fermion 
plus a five-dimensional  bosonic spinor field with the action defined by the 5-dim. 
Wilson-Dirac operator square $\vert D_{\rm w (5dim.)}-m_0 \vert^2$, 
both subject to the anti-periodic condition in the fifth-direction. ($m_0$ is the domain wall height in the lattice unit.)
Then one can safely take the limit of the infinite extent of the fifth dimension 
and the reflection positivity of overlap Dirac fermions follows indeed. 

In the case with gauge interaction, however,  
the positivity condition is not satisfied for the Pauli-Villars bosonic field. 
In fact, 
for the action of  the Pauli-Villars bosonic field, 
\begin{equation}
A_{\rm PV (b)} =- \sum_x   \phi^\ast(x)  \vert D_{\rm w (5dim.)}-m_0 \vert^2 \phi(x) , 
\end{equation}
one has
\begin{equation}
\theta (A_{\rm PV (b)} ) \not = A_{\rm PV (b)} 
\end{equation}
under the anti-linear $\theta$ operation 
for the bosonic spinor field $\phi_\alpha(x)$ and the link variable $U(x,y)$, 
\begin{equation}
 \theta \phi_\alpha(x) : = \phi_\alpha (\theta x), \quad \theta U(x,y)=U(\theta x,\theta y), 
\end{equation}
and for an observable $F$ of the form $F(U,\phi,\psi)=f(\phi,U)M(\psi)$,
\begin{align}
\theta(F)(U,\phi,\psi)=f^*(\theta\phi,\theta U)M^\dagger(\theta\psi), 
\end{align} 
as before.
This is 
due to the non-vanishing commutators of 
the covariant difference operators $ [ \nabla_0, \nabla_k] \not = 0$ $(k=1,2,3)$.  
And this causes a difficulty in completing the proof.
Of course,  it does not exclude the possibility that 
the overlap fermion itself, which is defined in the limit of the infinite extent of the fifth dimension, 
satisfies the reflection positivity.
In this approach with domain wall fermion,  it would be possible to trace the effects of the violation of 
the reflection positivity in the limit of the infinite extent of the fifth dimension. 
Work in this direction is in progress. 

\vspace{.5em}
\noindent
\section*{Acknowledgements}
K.U. would like to thank Tsutomu T. Yanagida for valuable encouragements.
K.U. is supported by Global COE Program ``the Physical Science Frontier", MEXT, Japan.
This work was supported by World Premier International Center Initiative (WPI Program), MEXT, Japan.
Y.K. would like to thank S.~Hashimoto and H.~Fukaya for discussions.
Y.K. is supported in part by Grant-in-Aid for Scientific Research No.~21540258, ~21105503.


\begin{thebibliography}{99}

\bibitem{Karsten:1980wd}
  L.~H.~Karsten and J.~Smit,
  Nucl.\ Phys.\  B {\bf 183}, 103 (1981).

\bibitem{Nielsen:1980rz}
  H.~B.~Nielsen and M.~Ninomiya,
  Nucl.\ Phys.\  B {\bf 185}, 20 (1981)
  [Erratum-ibid.\  B {\bf 195}, 541 (1982)].
  
\bibitem{Nielsen:1981xu}
  H.~B.~Nielsen and M.~Ninomiya,
  Nucl.\ Phys.\  B {\bf 193}, 173 (1981).


\bibitem{Ginsparg:1981bj}
  P.~H.~Ginsparg and K.~G.~Wilson,
  Phys.\ Rev.\  D {\bf 25}, 2649 (1982).

\bibitem{Neuberger:1997fp}
H.~Neuberger,
Phys.\ Lett.\ B {\bf 417}, 141 (1998) .

\bibitem{Neuberger:1998wv}
  H.~Neuberger,
  Phys.\ Lett.\  B {\bf 427}, 353 (1998) . 

\bibitem{Hasenfratz:1998ri}
P.~Hasenfratz, V.~Laliena and F.~Niedermayer,
Phys.\ Lett.\ B {\bf 427}, 125 (1998) . 

\bibitem{Hasenfratz:1998jp}
P.~Hasenfratz,
Nucl.\ Phys.\ B {\bf 525}, 401 (1998) .


  
\bibitem{Luscher:1998pqa}
  M.~Luscher,
  Phys.\ Lett.\  B {\bf 428}, 342 (1998) . 

\bibitem{Kaplan:1992bt}
  D.~B.~Kaplan,
  Phys.\ Lett.\  B {\bf 288}, 342 (1992) . 

\bibitem{Shamir:1993zy}
  Y.~Shamir,
  Nucl.\ Phys.\  B {\bf 406}, 90 (1993) . 
  
\bibitem{Furman:1994ky}
  V.~Furman and Y.~Shamir,
  Nucl.\ Phys.\  B {\bf 439}, 54 (1995) . 
  
  
  
\bibitem{Hernandez:1998et}
  P.~Hernandez, K.~Jansen and M.~Luscher,
  Nucl.\ Phys.\  B {\bf 552}, 363 (1999) . 
  
\bibitem{Fukaya:2007fb}
  H.~Fukaya {\it et al.}  [JLQCD Collaboration],
  Phys.\ Rev.\ Lett.\  {\bf 98}, 172001 (2007) . 
 
\bibitem{Fukaya:2009fh}
  H.~Fukaya, S.~Aoki, S.~Hashimoto, T.~Kaneko, J.~Noaki, T.~Onogi and N.~Yamada
  Phys.\ Rev.\ Lett.\  {\bf 104}, 122002 (2010) . 

\bibitem{Hashimoto:2008fc}
  S.~Hashimoto,
  PoS {\bf LATTICE2008}, 011 (2008), and references there in.
  
\bibitem{Luscher:2000hn}
  M.~Luscher,
  arXiv:hep-th/0102028.

\bibitem{Creutz:2004ir}
  M.~Creutz,
  Phys.\ Rev.\  D {\bf 70}, 091501 (2004) . 

\bibitem{Mandula:2009yd}
  J.~E.~Mandula,
  Phys.\ Rev.\  D {\bf 80}, 085023 (2009) . 
  
  
\bibitem{Osterwalder:1977pc}
  K.~Osterwalder and E.~Seiler,
  Annals Phys.\  {\bf 110}, 440 (1978).
  
\bibitem{Luscher:1976ms}
  M.~Luscher,
  Commun.\ Math.\ Phys.\  {\bf 54}, 283 (1977).
  
\bibitem{Menotti:1987cq}
  P.~Menotti and A.~Pelissetto,
  Nucl.\ Phys.\ Proc.\ Suppl.\  {\bf 4}, 644 (1988).

\cite{Golterman:2003qe}
\bibitem{Golterman:2003qe}
  M.~Golterman and Y.~Shamir,
  Phys.\ Rev.\  D {\bf 68}, 074501 (2003)
  [arXiv:hep-lat/0306002].


\bibitem{Osterwalder:1973dx}
  K.~Osterwalder and R.~Schrader,
  Commun.\ Math.\ Phys.\  {\bf 31}, 83 (1973).
  
\bibitem{Osterwalder:1974tc}
  K.~Osterwalder and R.~Schrader,
  Commun.\ Math.\ Phys.\  {\bf 42}, 281 (1975).
  
\bibitem{Luscher:1984is}
  M.~Luscher and P.~Weisz,
  Nucl.\ Phys.\  B {\bf 240}, 349 (1984).


\bibitem{Horvath:1998cm}
  I.~Horvath,
  Phys.\ Rev.\ Lett.\  {\bf 81}, 4063 (1998) . 

\bibitem{Neuberger:1997bg}
  H.~Neuberger,
  Phys.\ Rev.\  D {\bf 57}, 5417 (1998) . 
 
\bibitem{Kaplan:1999jn}
  D.~B.~Kaplan and M.~Schmaltz,
  Chin.\ J.\ Phys.\  {\bf 38}, 543 (2000) . 
 
%
  
\bibitem{Narayanan:1993sk}
  R.~Narayanan and H.~Neuberger,
  Nucl.\ Phys.\  B {\bf 412}, 574 (1994) . 
\bibitem{Narayanan:1993ss}
  R.~Narayanan and H.~Neuberger,
  Phys.\ Rev.\ Lett.\  {\bf 71}, 3251 (1993). 

\bibitem{Narayanan:1994gw}
  R.~Narayanan and H.~Neuberger,
  Nucl.\ Phys.\  B {\bf 443}, 305 (1995). 
\bibitem{Luscher:1999mt}
  M.~Luscher,
  Nucl.\ Phys.\ Proc.\ Suppl.\  {\bf 83}, 34 (2000). 
  
%
\bibitem{Kikukawa:1999sy}
  Y.~Kikukawa and T.~Noguchi,
  arXiv:hep-lat/9902022.
%
%

\end{thebibliography}
\end{document}